======================================================================== 749
\documentstyle[12pt]{article}
\begin{document}

\def\beq{\begin{equation}}
\def\eeq{\end{equation}}
\def\bce{\begin{center}}
\def\ece{\end{center}}
\def\bea{\begin{eqnarray}}
\def\eea{\end{eqnarray}}
\def\ben{\begin{enumerate}}
\def\een{\end{enumerate}}
\def\ul{\underline}
\def\ni{\noindent}
\def\nn{\nonumber}
\def\bs{\bigskip}
\def\ms{\medskip}
\def\wt{\widetilde}
\def\tr{\mbox{Tr}\, }
\def\brr{\begin{array}}
\def\err{\end{array}}
\def\dsp{\displaystyle}

\hfill UB-ECM-PF 95/2

\hfill February 1995

\vspace*{3mm}

\begin{center}

{\LARGE \bf
On black holes in the theory of dilatonic gravity coupled to a
scalar
field}

\vspace{1cm}

{\bf E. Elizalde$^{a,b,}$}\footnote{E-mail:
eli@zeta.ecm.ub.es},
{\bf P. Fosalba-Vela$^{a}$},
{\bf S. Naftulin$^{c}$}
and {\bf S.D. Odintsov$^{b,}$}\footnote{On
 leave of absence from Tomsk Pedagogical Institute,
634041 Tomsk, Russia. E-mail: odintsov@ecm.ub.es}
\vspace{3mm}

$^a$Center for Advanced Study CEAB, CSIC,
Cam\'{\i} de Sta. B\`arbara, 17300 Blanes \\
$^b$Department ECM and IFAE,
Faculty of Physics, University of  Barcelona, \\
Diagonal 647, 08028 Barcelona,
Catalonia, Spain \\
$^c$Institute for Single Cristals, 60 Lenin Ave., 310141 Kharkov,
Ukraine \\

\vspace{15mm}

{\bf Abstract}
\end{center}

Taking advantage of the representation of dilatonic gravity with
the
$R^2$-term under the form of low-derivative dilatonic gravity
coupled
to an additonal scalar, we construct a general renormalizable model
motivated by
this theory. Exact black hole solutions are found for some
specific
versions of the model, and their thermodynamical properties are
described in detail. In particular, their horizons and
temperatures are
calculated. Finally, the corresponding one-loop effective action
is
obtained in the conformal gauge, and a number of its properties
---including the construction of one-loop finite models---
are briefly described.

\vfill


\newpage

\ni{\bf 1. Introduction.}
Recently, the study of two-dimensional (2D) dilatonic gravity has
become
very interesting,  both at the classical and at the quantum
level, for a
variety of reasons. In the first place, this theory is closely
connected
with string theory (see \cite{15} for a review on this point),
where it
appears as a sort of effective action. Secondly, dilatonic
gravity
itself can be represented under the form of a sigma model (see,
for
example, Ref. \cite{16}), what helps to understand the connection
between the conformal properties of the corresponding sigma model
and
the solutions of dilatonic gravity \cite{14}. Finally, the hope
exists
that through the study of `easier' 2D models one can get some
insights
useful for the investigation of realistic 4D gravity.

So far, the main activity has been concentrated on the study of ordinary
dilatonic gravity with matter. However, there are different
motivations for studying in such theory ---in addition to the
dilaton---
another scalar of similar nature. For example, in string theory
this
scalar is the modulus field, which is connected with the radius
of the
compactified space (see, for example, \cite{17}). In charged
string
theory there also appears an additional scalar, called spectator
field
(see, for instance, \cite{18}). The additional scalar may be
interpreted
as a Liouville field \cite{6}, in some cases. Furthermore, as we
will
show in the next section, the scalar field appears in dilatonic
gravity
with an $R^2$-term, on lowering the number of derivatives.

In this letter we formulate the theory of renormalizable
dilatonic
gravity with matter, coupled to a scalar field (Sect. 2). The
study of
solutions of black hole type for a few different variants of such
theory
is carried out in Sect. 3, with a systematic description of their
properties. Finally, Sect. 4 is devoted to the study of the one-loop
effective action of the model under consideration.
\ms

\ni{\bf 2. Dilatonic gravity coupled to a scalar field.}
We start from the theory of dilatonic gravity with an $R^2$-term.
The
Lagrangian can be written in the following form
 \begin{equation}
L = \frac{1}{2} Z(\phi) g^{\mu\nu}
\partial_\mu \phi\partial_\nu \phi +
C(\phi) R+ V(\phi) + \omega (\phi) R^2
- {1 \over 2} f(\phi) g^{\mu\nu} \partial_\mu \chi_i
\partial_\nu \chi^i,
 \label{521}
\end{equation}
where $\phi$ is the dilaton field, $\chi_i$ are scalars ($i=1,2,
\ldots, n$), and where the dilatonic functions $Z(\phi)$,
$C(\phi)$,
$V(\phi)$, $\omega (\phi)$  and $f(\phi)$ are assumed to be
analytic in
$\phi$.
At the classical level, the theory (\ref{521}) represents a
particular
case of the most general theory of higher-derivative dilatonic
gravity
that was studied in Ref. \cite{1}. The theory is motivated by the
consideration of string theory in the background of massive modes
(for a
recent discussion see, for instance, \cite{2}). At the quantum
level,
however, the theory (\ref{521}) belongs to a different class than
the
models of \cite{1}. In particular, one is not able to obtain the
one-loop effective action of theory (\ref{521}) from the general
one-loop effective action in \cite{1} (since the  corresponding
limit is
singular).

As is well known, the calculations in the theory of
higher-derivative
quantum gravity are very involved (for a review, see \cite{3})
and it is
always preferable to work, when possible, with a low-derivative
theory.
One can reduce the order of the derivatives in (\ref{521}) by
introducing
an auxiliary scalar, $\psi =2 \omega (\phi ) R$, according to the
trick
in Ref. \cite{4}:
 \bea
S &=& \int d^2x \, \sqrt{-g} \left\{ \frac{1}{2} Z(\phi)
 g^{\mu\nu} \partial_\mu \phi\partial_\nu \phi + [
C(\phi) + \psi ] R - \frac{1}{4\omega (\psi)} \psi^2+ V(\phi) \nn
\right. \\ && \left.
 - {1 \over 2} f(\phi) g^{\mu\nu} \partial_\mu \chi_i
\partial_\nu \chi^i \right\}.
 \label{522}
\eea
This new theory is equivalent to the theory (\ref{521}). At the same
time, we
get in this way an example of dilatonic gravity coupled with an
additional scalar, which can be also interpreted as a Liouville
field or
 a matter scalar, or as a kind of spectator field ---as we have
mentioned in the introduction.

For what has been said, it is interesting to spend some time in
the
construction of the most general theory of renormalizable
dilatonic
gravity  coupled to an additional scalar field. Starting
from (\ref{521}), its natural generalization looks like
 \bea
S &=& \int d^2x \, \sqrt{-g} \left\{ \frac{1}{2} Z_{(ij)}(\phi)
g^{\mu\nu}
\partial_\mu \phi^i\partial_\nu \phi^j +
C(\phi)  R + V(\phi) \right. \nn \\ && \left.
 - {1 \over 2} f(\phi) g^{\mu\nu} \partial_\mu \chi_a
\partial_\nu \chi^a\right\},
 \label{523}
\eea
where we have now two scalars $\phi_i$, $i=1,2$,
representing
the dilaton and Liouville (or matter) scalar field, while the
$\chi_a$
are the matter scalars. The kinetic matrix $Z_{(ij)} (\phi)$ is
considered
to be symmetric and the functions $C(\phi )$, $V(\phi )$ and
$f(\phi)$
depend now (obviously) on both fields $\phi_1$ and $\phi_2$. It
is
easy to see that the theory (\ref{523}) is renormalizable in a
generalized sense
(unlike Einstein's 2D gravity \cite{19}). Its one-loop
renormalizability
will be discussed in Sect. 4.
\ms

\ni{\bf 3. Solutions of black hole type.}
To begin with, we will consider some simple versions of the
theory
(\ref{521}) (without matter), at the classical level, and will
search for solutions of black hole type. In all we will consider
four
different models, each of them being a particular representative
of
(\ref{523}), given by the Lagrangians:
\bea
L_I &=& \frac{1}{8\pi G} \psi (R+\Lambda_1) + b
 g^{\mu\nu} \partial_\mu \phi\partial_\nu \phi + \gamma \phi R +
e^{-2a\phi} \Lambda, \label{524} \\
L_{II} &=& \frac{1}{8\pi G} \left(  \psi R -
\frac{\psi^2}{4\omega(\phi)} \right) + b
 g^{\mu\nu} \partial_\mu \phi\partial_\nu \phi + \gamma \phi R +
e^{-2a\phi} \Lambda, \label{525} \\
L_{III} &=& \frac{1}{8\pi G} \psi (R+\Lambda_1) +
 e^{a_1 \phi}
 g^{\mu\nu} \partial_\mu \phi\partial_\nu \phi +
e^{-2a\phi} \Lambda +
\gamma e^{a_2\phi} R, \label{526} \\
L_{IV} &=& \frac{1}{8\pi G} \left(  \psi R -
\frac{\psi^2}{4\omega(\phi)} \right) +
 e^{a_1 \phi}
 g^{\mu\nu} \partial_\mu \phi\partial_\nu \phi +
e^{-2a\phi} \Lambda +
\gamma e^{a_2\phi} R. \label{527}
\eea
The first Lagrangian, $L_I$, describes  Jackiw-Teitelboim (JT)
dilatonic
gravity \cite{5}, with the dilaton $\psi$ interacting with a
Liouville
theory ($\phi$ is the Liouville field). Model II corresponds to
the
interaction of some different dilatonic gravity with a Liouville
theory.
This model can be also interpreted (after eliminating $\psi$ and
rescaling $\gamma \rightarrow \gamma/(8\pi G)$) as a particular
version
of dilaton gravity with an $R^2$-term (\ref{521}). R.B. Mann has
recently considered black hole type solutions in models similar
to
(\ref{524}) and (\ref{525}) (albeit with a different dilatonic
gravity
part, see \cite{6,8}). Here we will use a technique similar to
that of
Ref. \cite{6}, specially to calculate the ADM mass (in the limit
$x\rightarrow \infty$). Model III represents the interaction of
JT
gravity with a bosonic string-like effective action. Finally, the
Lagrangian IV describes another version of the dilatonic gravity
(\ref{521}) with an $R^2$-term (again with
 $\gamma \rightarrow \gamma/(8\pi G)$).

Choosing a static metric of the form
\beq
ds^2 =-g(x) dt^2 + g(x)^{-1} dx^2
\label{528}
\eeq
one can solve the classical field equations and try to  find
configurations corresponding to black holes. The results are
given in Tables 1 to 4. Two general comments are in order.
 Concerning the $\pm$ sign of
the horizon, it is understood that when there is only one horizon
the
$+$ sign is chosen. Moreover, we always take $x=|x| =r$, so that
$x$ is
assumed to be positive.

\begin{table}
\vspace*{-15mm}

\begin{center}
\begin{tabular}{|c|c|}
\hline \hline
Characteristics  & Model I \\
 \hline \hline  Fields & $g(x) = \dsp \frac{a^2\Lambda}{b} +Dx +
\dsp \frac{\Lambda_1}{2}x^2, \ \ \psi (x) = Bx, \ \ \phi (x) =
\frac{1}{a}
\ln x $ \\
\hline  Parameters & $D = \dsp\frac{4\pi Gb}{a^2B} \Lambda_1$, \ \
$B\neq 0$
arbitrary \\
 \hline {\cal M} (ADM mass)  & $-\dsp \frac{2\pi Gb^2}{a^4B} \Lambda_1
+
\dsp \frac{Ba^2}{8\pi Gb}\Lambda$ \\
 \hline  BH criteria & (i) \ \  $\Lambda_1 >0 \ \ \ \ (B<0$ always)\\
\hline  two horizons  & if  $\Lambda >0, \ \  b >0, \ \ b^3 >
\dsp\frac{a^6B^2\Lambda}{8\pi^2G^2\Lambda_1}$ \\
\hline  one horizon  & if  $\Lambda \cdot b <0$ \\
 \hline  BH criteria & (ii) \ \  $\Lambda_1 <0 \ \ \ \ (B>0$ always)\\
\hline  two horizons  & if  $\Lambda >0, \ \ b <0 \ \ |b^3| >
\dsp\frac{a^6B^2\Lambda}{8\pi^2G^2|\Lambda_1|}$ \\
\hline  one horizon  & if  $\Lambda \cdot b >0$ \\
\hline Curvature & $-\Lambda_1$ \\
\hline Horizon & $-\dsp\frac{4\pi Gb}{a^2B} \pm \sqrt{\left(
\dsp\frac{4\pi Gb}{a^2B} \right)^2 - \dsp\frac{2a^2\Lambda}{\Lambda_1B}}$
\\
\hline Temperature & $\dsp\frac{|\Lambda_1|}{4\pi}  \sqrt{\left(
\dsp\frac{4\pi Gb}{a^2B} \right)^2 - \dsp\frac{2a^2\Lambda}{\Lambda_1B}}$
\\
\hline \hline \end{tabular}

\caption{{\protect\small Characteristics of the solutions of
black hole type corresponding to model I.} \label{52t1}}

\end{center}

\end{table}

\begin{table}
\vspace*{-1cm}

\begin{center}
\begin{tabular}{|c|c|}
\hline \hline
Characteristics  & Model II \\
 \hline \hline  Fields & $g(x) = \dsp\frac{a^2\Lambda}{b} +Ax +
C x^2, \ \ \psi (x) = -\dsp\frac{8\pi Gb}{a^2}, \ \ \phi (x) =
\dsp\frac{1}{a}
\ln x $ \\
\hline  Parameters & $C = \dsp\frac{2\pi Gb}{a^2\omega_0}$, \ \
$A\neq 0$ arbitrary, \ \ $\omega (\phi) =\omega_0=$ const. \\
 \hline {\cal M} (ADM mass)  & $- \dsp\frac{bA}{2a^2} $\\
 \hline  BH criteria & (i) \ \  $b/\omega_0 >0$ \\
\hline  two horizons  & if  $\Lambda >0, \ \  A <0, \ \ A^2 >
\dsp\frac{8\pi G\Lambda}{\omega_0}$ \\
\hline  one horizon  & if  $A \cdot \Lambda >0$ \\
 \hline  BH criteria & (ii) \ \  $b/\omega_0 <0$ \\
\hline  two horizons  & if  $\Lambda <0, \ \  A <0, \ \ A^2 >
\dsp\frac{8\pi G\Lambda}{\omega_0}$ \\
\hline  one horizon  & if  $A \cdot \Lambda <0$
 \\ \hline Curvature & $-\dsp\frac{4\pi Gb}{a^2\omega_0}$ \\
\hline Horizon & $\dsp\frac{\omega_0 |A|}{4\pi Gb} \left(-1 \pm
\sqrt{1- \dsp\frac{8\pi G \Lambda}{\omega_0A^2}}\right)$ \\
\hline Temperature & $\dsp\frac{|A|}{4\pi}  \sqrt{ 1
 - \dsp\frac{8\pi G\Lambda}{\omega_0A^2}}$ \\
\hline \hline \end{tabular}

\caption{{\protect\small Characteristics of the solutions of
black hole type corresponding to model II.} \label{52t2}}

\end{center}

\end{table}

\begin{table}
\begin{center}
\begin{tabular}{|c|c|}
\hline \hline
Characteristics  & Model III \\
 \hline \hline  Fields & $g(x) = C +Dx +
\dsp\frac{\Lambda_1}{2} x^2, \ \ \psi (x) = \dsp\frac{A}{x^2} +
\dsp\frac{B}{x^4}, \
\ \phi (x) = E \ln x $ \\
\hline  Parameters & $a_1= \dsp\frac{a_2}{2} =-\dsp\frac{4a}{3}$,
$A = \dsp\frac{64\pi G}{9a_2^2} \, \dsp\frac{48\gamma + 9a_2^2
(\Lambda/\Lambda_1)^2}{16 \gamma + 3a_2^2
(\Lambda/\Lambda_1)^2}$,
  $B=-8\pi G\gamma$,
 \\  &
  $C= -\dsp\frac{\gamma a_2^2\Lambda_1}{16}$,
  $D= -\dsp\frac{3a_2^2\Lambda}{32}$, $E=-\dsp\frac{4}{a_2}$   \\
 \hline  BH criteria & (i) \ \  $\Lambda_1  >0$ \\
\hline  two horizons  & if  $\Lambda <0, \ \  \gamma <0, \ \
a_2^2 >
\dsp\frac{128\Lambda_1^2}{9\Lambda^2} |\gamma |$ \\
\hline  one horizon  & if  $\gamma >0$ \\
 \hline  BH criteria & (ii) \ \  $\Lambda_1  <0$ \\
\hline  two horizons  & if  $\Lambda >0, \ \  \gamma <0, \ \
a_2^2 >
\dsp\frac{128\Lambda_1^2}{9\Lambda^2} |\gamma |$ \\
\hline  one horizon  & if  $\gamma >0$ \\
  \hline Curvature & $-\Lambda_1$ \\
\hline Horizon & $\dsp\frac{3a_2^2\Lambda}{32\Lambda_1} \left(-1 \pm
\sqrt{1+ \dsp\frac{128\gamma \Lambda^2}{9a_2^2\Lambda_1^2}}\right)$
\\
\hline Temperature & $\dsp\frac{|\Lambda_1a_2|}{16\pi}  \sqrt{
  \dsp\frac{9a_2^2\Lambda^2}{64\Lambda_1^2} +2\gamma}$ \\
\hline \hline \end{tabular}

\caption{{\protect\small Characteristics of the solutions of
black hole type corresponding to model III.} \label{52t3}}

\end{center}

\end{table}

\begin{table}
\begin{center}
\begin{tabular}{|c|c|}
\hline \hline
Characteristics  & Model IV \\
 \hline \hline  Fields & $g(x) = \dsp\frac{A}{x} +Bx,
  \psi (x) = -4A \left( \dsp\frac{C}{x^5} + \dsp\frac{E}{x^7} \right),
 \phi (x) = D \ln x $, \\   & $\omega (\phi ) = C e^{-2\phi /D} +
E e^{-4\phi
/D} = \dsp\frac{C}{x^2}+\dsp\frac{E}{x^4}$ \\
\hline  Parameters & $a_1=\dsp\frac{7a_2}{5}
=-\dsp\frac{7a}{5}$, $A = \dsp\frac{28a_2^2\Lambda}{775}$,
  $B=-\dsp\frac{224 a_2^4\Lambda \gamma}{135625}$,
 \\  &
 $C=
\dsp\frac{1550 \pi G\gamma}{28 a_2^2\Lambda}$,
  $D= -\dsp\frac{5}{a_2}$, $E=-\dsp\frac{19375 \pi G}{784 a_2^4\Lambda}$
\\
 \hline  BH criteria &  $\gamma >0$, $\Lambda <0$ \\
\hline  one horizon  & always \\
  \hline Curvature & $-\dsp\frac{56 a_2^2\Lambda}{775 x}$ \\
\hline Horizon & $\sqrt{\dsp\frac{175}{8a_2^2\gamma}}$ \\
\hline Temperature & $\dsp\frac{112 a_2^4\gamma |\Lambda|}{135625\pi}$
\\
\hline \hline \end{tabular}

\caption{{\protect\small Characteristics of the solutions of
black hole type corresponding to model IV.} \label{52t4}}

\end{center}

\end{table}

For the first three models we have found cosmological black hole
type solutions, which have been analyzed both for asymptotically
de Sitter and anti-de Sitter spacetimes. As we see from the
tables, a wide variety of cases with single and double horizons
is obtained, depending on the sign of the parameters.

The $\Lambda =0$ case for models I and II can be obtained
strightforwardly by just taking the $\Lambda \rightarrow 0$ limit
of all quantities shown on the  first two tables, except for the
black hole criteria since, in that case, we find one horizon
only, located at $x_H =- 8\pi Gb /(a^2B)$, with $\Lambda, b >0$
and $B<0$, for the first model, and at  $x_H =- \omega_0 a^2A
/(2\pi Gb)$, with $b, \omega_0 >0$ and $A<0$, for the second one.

The solution for model IV corresponds to a real black hole with an
event horizon. It exhibits a metric and curvature singularity at
$x=0$. Notice that for $\gamma <0$ this singularity will be a
naked one. Other interesting quantities, as the ADM mass (which turns
out to be zero or, better, `compatible' with zero), the curvature and
the temperature of the black holes are also
given, in order to obtain a sufficiently detailed picture of the models
considered.

Summing up, we have investigated here the black hole solutions of
several models of dilatonic gravity coupled to a Liouville theory.
It is interesting to observe that in the case of JT dilatonic
gravity (without the field $\phi$) a regular black hole of
cosmological type has been first discovered in Ref. \cite{7}. Our
results provide a straightforward extension of this type of black holes
for the case when an additional scalar field is present. In a similar
way one can
construct black hole solutions for the other versions of (\ref{523}).
\ms

\ni{\bf 4. The one-loop effective action.}
Let us now consider the theory (\ref{523}) at the quantum level.
Using the standard Schwinger-De Witt algorithm and the background
field method, the corresponding quantum effective action can be
calculated without problems. For the case of standard dilatonic
gravity this has been done in great detail in Refs.
\cite{9,10}. That calculation can be repeated here, to
deal with the more complicated cases in which we are interested.
The starting point of the background field method is now
\beq
\phi_i \longrightarrow \phi_i + \varphi_i, \ \ \chi_a
\longrightarrow \chi_a + \eta_a, \ \ g_{\mu\nu} \longrightarrow
e^\sigma g_{\mu\nu}.
\label{529}
\eeq
The calculation is, however, rather tedious and straightforward,
since it just makes use of the well-known techniques mentioned
already, so we have decided to spare the reader all details and to
present the final results only. They correspond to the one-loop
effective action in the conformal gauge, and are:
\bea
\Gamma_{div} &=& \frac{1}{4\pi \epsilon} \int d^2x\, \sqrt{-g}
\left\{ \left( \frac{23-n}{6} + C_i^i\right) R + V_i^i + 2 G_i^*
V^i + G^{**} V \right. \nn\\ && + \left( \frac{f_i^i}{2} -
\frac{f_if^i}{2f} \right)  g^{\mu\nu} \partial_\mu \chi_a
\partial_\nu \chi^a - \left( Z_{(ij)}^i+ G^{i*}C_{ij} \right) \Delta
\phi^j \nn \\ && \left. + \left[ \frac{1}{4} A_{ij} (\phi) +n
\frac{f_if_j}{4f^2} \right]  g^{\mu\nu} \partial_\mu \phi_i
\partial_\nu \phi_j \right\},
 \label{529p}
\eea
where
\beq
G_{AB} = \left( \brr{cc} Z_{(ij)} & C_j \\ C_i & 0 \err \right), \
\ \ \ G^{AB} = (G_{AB})^{-1} =\left( \brr{cc} G^{ij} & G^{j*} \\
G^{i*} & G^{**} \err \right), \label{5210}
\eeq
and $G_{AB}$ is supposed to be invertible. The indices are rised
with the spacetime metric corresponding to the inverse
configuration, $G^{ij}$, for instance,
\beq
V^i = G^{ij} V_j, \qquad \qquad Z^{(i}_{\ j)} = G^{ik}Z_{(kj)}.
\eeq
The derivatives of the functions in (\ref{529p}) are denoted with
lower latin indices, e.g. $C_i = \frac{\delta C}{\delta \phi_i}$
and, in  (\ref{529p}),
\bea
{\cal A}_{kl} (\phi) &=& Z^{(ij)}_k Z_{(ij)l} + 2Z_{(ik)}^j
Z_{(jl)}^i - 2 Z^{(i}_{ \ k)j} Z_{\ (il)}^j + 4G^{i*}G^{j*}C_{ik}
C_{jl} \nn \\ &&+ 2\left[ G^{i*}C^j_k (Z_{(ij)l} + Z_{(jl)i} -
Z_{(il)j} ) + (k \leftrightarrow l ) \right]
\nn \\ &&+ 2Z_{(kl)i}^i -2 Z_{(ik)l}^i -2Z_{(il)k}^i-4 G^{i*}
C_{ikl}. \label{5212}
\eea

The one-loop effective action (\ref{529p}) looks rather
complicated, although it shows explicitly that the theory is
one-loop renormalizable. By looking at the zeros of the
generalized beta functions one should be able to find the fixed
points of the generalized renormalization group, and to construct
the corresponding finite dilatonic gravity models of type
(\ref{523}).
In particular, it follows from (\ref{529p}) that
the theory (\ref{523}) with all dilatonic coupling functions
being constants is finite (for $V=0$, $n=23$), and such a model is
certainly a fixed point of the renormalization group.

Specifying the general answer (\ref{5210}) to the model (\ref{522}) one
can easily obtain the one-loop effective action corresponding to
(\ref{522}), e.g. in the formulation of the theory with the auxiliary
field $\psi$. After eliminating the auxiliary field by means of the
constraint $\psi = 2\omega (\phi) R$ ---the usual way to proceed
in the background field method (see, for example, \cite{20})--- we get
\bea
\Gamma_{div} &=& \frac{1}{4\pi \epsilon} \int d^2x\, \sqrt{-g}
\left\{ \left( \frac{11-n}{6} + \frac{C''}{Z} -\frac{2C'\omega'}{\omega
Z} \right) R - \frac{(1/\omega)''\omega^2}{Z} R^2 + \frac{V''}{Z}-
\frac{(C')^2}{2\omega Z}  \right. \nn\\ && + \left. \left(
\frac{n(f')^2}{4f^2}
- \frac{3(Z')^2}{4Z^2} + \frac{Z''}{2Z} \right)  g^{\mu\nu} \partial_\mu
\phi \partial_\nu \phi - \left( \frac{f''}{2Z}- \frac{(f')^2}{2fZ}
 \right)
  g^{\mu\nu} \partial_\mu \chi_i \partial_\nu \chi^i \right\},
 \label{52n1}
\eea
The $R^2$-term in (\ref{52n1}) can be eliminated via the equation of
motion $\delta S /\delta \phi =0$ for the theory (\ref{521}).

Starting from (\ref{521}) a whole set of models that are one-loop finite
can be given, which hence correspond to the fixed points of the
generalized renormalization group. One of this examples is given by
model (\ref{521}) itself with the following dilatonic functions:
\bea
&& n=11, \ \ \ \ Z(\phi ) = \frac{\alpha_3 \exp (\alpha_2 \phi
)}{[\alpha_4 \exp (2\alpha_2 \phi) +1 ]^2}, \ \ \ \ f(\phi ) = \alpha_1
\exp (\alpha_2 \phi /\sqrt{11}), \nn \\
&& \omega (\phi )= \mbox{const.}, \ \ \ \ C(\phi )= 2\alpha_5 \phi +
\alpha_6, \ \ \ \ V(\phi ) = \frac{\alpha_5^2}{\omega} \phi^2 + \alpha_7
\phi + \alpha_8,
 \label{52n2}
\eea
where $\alpha_1, \ldots, \alpha_8$, are arbitrary constants.

It is also interesting to remark that ---as in the case of
standard dilatonic gravity (see \cite{11} and the paper by
Chamseddine in \cite{5})--- the model (\ref{523})
can be easily represented, in the conformal gauge $g_{\mu\nu} =
e^\sigma \bar{g}_{\mu\nu}$, as a sigma model of the special form:
\beq
S= \int d^2 x \, \sqrt{-\bar{g}} \left( \frac{1}{2}
G_{\tilde{A}\tilde{B}} \bar{g}^{\mu\nu} \partial_\mu
\tilde{\phi}^{\tilde{A}}  \partial_\nu \tilde{\phi}^{\tilde{B}}
+\bar{R} \psi + T \right),
\label{5213}
\eeq
with $\tilde{\phi}^{\tilde{A}} = (\phi_i, \sigma, \chi_a)$, and
where
\beq
G_{\tilde{A}\tilde{B}} = \left( \brr{ccc} Z_{(ij)} & C_j & 0 \\
C_i & 0 & 0 \\ 0 & 0 & f(\phi ) \delta_{ab} \err \right), \ \ \
\psi = C( \phi ), \ \ \ T = e^\sigma V( \phi ). \label{5214}
\eeq
Then an alternative way to calculate the one-loop effective
action (\ref{5212}) is to use the standard string $\beta$-functions of
the $\sigma$-model approach \cite{12,13}, with the background
(\ref{5214}). An interesting (although not easy) task would be to
try to understand the connection between the conformal properties
of the $\sigma$-model (\ref{5213}) and those of the black hole
type solutions that appear in the related theory of
string-inspired gravity (similar to Ref. \cite{14}).
We plan to return to these questions
in the near future.

In summary, we have studied in this letter a quite general
 (one-loop) renormalizable theory of dilatonic gravity coupled to
a scalar field. For some particular cases of this theory, we have
investigated in some detail their cosmological black hole
solutions, obtaining their main characteristics, as the ADM mass,
the horizon structure and the black-hole temperature. To conclude, the
one-loop effective action has been obtained, and the precise
connection of the theory with its formulation in terms of a sigma
model has been described.
 \vspace{5mm}


\noindent{\large \bf Acknowledgments}

SDO would like to thank A. Chamseddine for discussions.
This work has been supported by DGICYT (Spain), project Nos.
PB93-0035 and SAB93-0024,  by CIRIT (Generalitat de Catalunya),
and by RFFR (Russia), project No. 94-02-03234.

\end{document}